\documentclass{svjour2}                    
\smartqed  
\usepackage{amsmath,amssymb,graphicx,epsfig}

\journalname{Foundations of Physics}
\journalname{Foundations of Physics}

\begin{document}

\title{A Note on Solid-State Maxwell Demon}

\titlerunning{Solid-State Maxwell Demon}

\author{Germano D'Abramo$^1$\\}

\authorrunning{Germano D'Abramo}

\institute{$^1$ Istituto Nazionale di Astrofisica, Via Fosso del 
Cavaliere 100, 00133, Roma, Italy\\
\email{Germano.Dabramo@iasf-roma.inaf.it}}

\date{Received: 24 November 2010 / Accepted: 22 October 2011}

\maketitle

\begin{abstract}

Since 2002, at least two kinds of laboratory-testable, solid-state 
Maxwell demons have been proposed that utilize the electric field energy 
of an open-gap {\em n-p} junction and that seem to challenge the 
validity of the Second Law of Thermodynamics. In the present paper we 
present some arguments against the alleged functioning of such devices.


\end{abstract}

\section{Introduction}

\begin{figure}[t]
\begin{center}
\includegraphics[height=10cm]{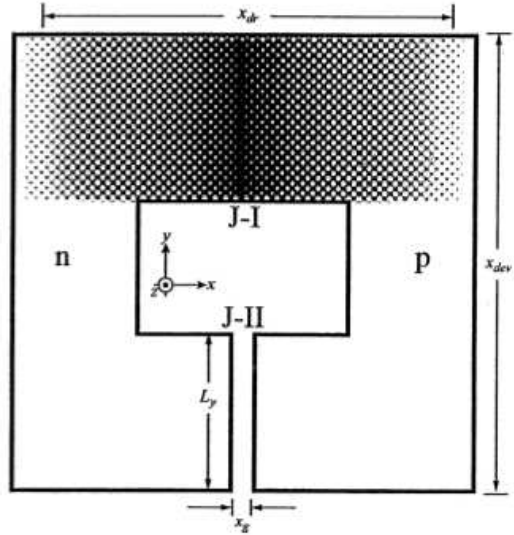}
\end{center}
\caption{The core of the solid-state Maxwell demon. This sketch is 
from Fig.~1 of reference~\cite{bib17a}.}
\label{fig1}
\end{figure}

Since 2002, two types of solid-state devices have been 
proposed~\cite{bib17a,bib17b,bib17c,bib19,cs} that basically utilize the 
electric field energy of an open-gap {\em n-p} junction and that seem to 
challenge the validity of the Second Law of Thermodynamics. They 
represent a sort of non-sentient solid-state Maxwell demons operating at 
room temperature, which are based on the cyclic electromechanical 
discharging and thermal recharging of the electrostatic potential energy 
intrinsic to the depletion region of a standard solid-state {\em n-p} 
junction. The core of their functioning is the shaped junction depicted 
in Fig.~\ref{fig1}.

It consists of two symmetric horseshoe-shaped pieces of {\em n}- and 
{\em p}-se\-mi\-con\-duc\-tor facing one another. At Junction I (J-I), 
the {\em n}- and {\em p}-regions are physically connected, while at 
Junction II (J-II) there is a vacuum gap whose width $x_g$ is small 
compared to the scale lengths of either the depletion region $x_{dr}$ or 
the overall device $x_{dev}$; namely, $x_g\ll x_{dr}\sim x_{dev}$. All 
the scale lengths are in the {\em micro-}, {\em nano-metric} realm.

As is well known from solid-state physics, a built-in potential $V_{bi}$ 
forms across the junction J-I, whose numerical value depends on the 
doping characteristics of the two regions (concentrations of donors and 
acceptors, intrinsic carrier concentration) and on the environmental 
temperature (in the present case, room temperature). Its value can be 
estimated analytically.

This potential is the result of charge diffusion across J-I as soon as 
the the two materials are physically joined. The depletion region is 
thus the region where, at equilibrium, {\em a balance between bulk 
electrostatic and diffusive (thermally driven) forces is attained}.
 
It is then claimed~\cite{bib17a,bib17b,bib17c,bib19,cs} that an electric 
field must exist also in J-II. According 
to~\cite{bib17a,bib17b,bib17c,bib19,cs}, the existence of an electric 
field in the J-II gap at equilibrium can be established either via 
Kirchhoff's loop rule (conservation of energy) or via Faraday's law 
($\oint \textrm{\bf E} \cdot d\textrm{\bf l} = 0$). In fact, with regard 
to the latter condition, it would be more proper to talk about {\em 
path-independence} of conservative electric fields rather than referring 
to Faraday's law. It is argued as follows. Consider a vectorial loop 
threading the J-I depletion region, the bulk of the device in 
Fig.~\ref{fig1}, and the J-II gap. Since the built-in electric field in 
the J-I depletion region is unidirectional, there must be a second 
electric field somewhere else along the loop to satisfy $\oint 
\textrm{\bf E} \cdot d\textrm{\bf l} = 0$. An electric field elsewhere 
in the semiconductor bulk (other than in the depletion region), however, 
would generate a current, which contradicts the assumption of 
equilibrium. Therefore, by exclusion, the other electric field must 
exist in the J-II gap. Kirchhoff's loop rule establishes the same 
result. Conservation of energy demands that a test charge conveyed 
around this closed path must undergo zero net potential drop; therefore, 
to balance $V_{bi}$ in the depletion region, there must be a 
counter-potential somewhere else in the loop. Since, at equilibrium, 
away from the depletion region in the bulk semiconductor there cannot be 
a potential drop (electric field) - otherwise there would be a 
non-equilibrium current flow, contradicting the assumption of 
equilibrium - the potential drop must occur outside the semiconductor; 
thus, it must be expressed across the vacuum gap J-II.

\begin{figure}[t]
\begin{center}
\includegraphics[height=11cm]{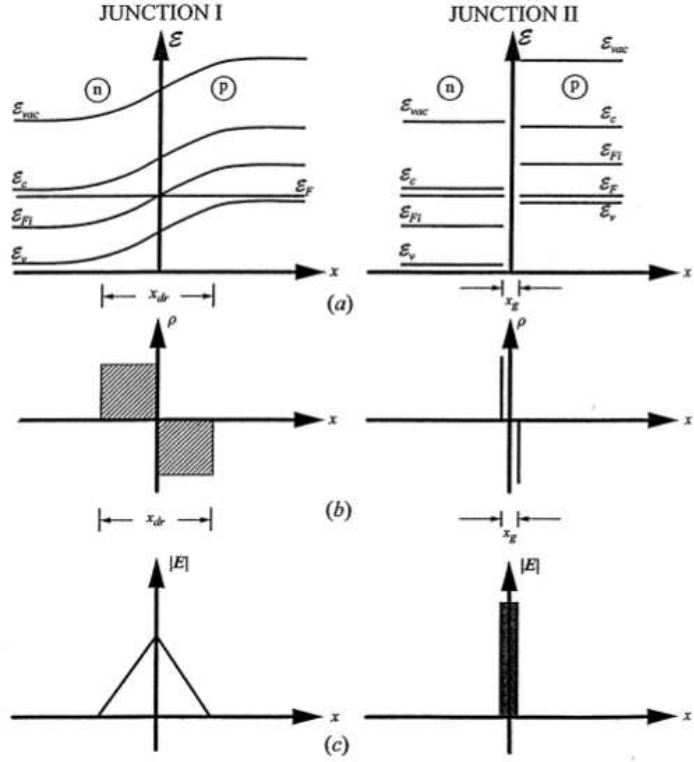}
\end{center}
\caption{Physical characteristics versus position $x$ through Junctions 
I and II. Left ($x < 0$) and right ($x > 0$) sides of each graph 
correspond to {\em n}- and {\em p}-regions, respectively. (a) Energy 
levels for vacuum (${\cal E}_{vac}$), conduction band edge (${\cal 
E}_c$), intrinsic Fermi level (${\cal E}_{Fi}$), Fermi level (${\cal 
E}_F$), valence band edge (${\cal E}_v$). (b) Charge density ($\rho$). 
(c) Electric field magnitude ($|\textrm{\bf E}|$). Note that vertical 
scales for $|\textrm{\bf E}|$ are different for J-I and J-II 
($|\textrm{\bf E}_{\textrm{\bf J-II}}|\gg |\textrm{\bf E}_{\textrm{\bf 
J-I}}|$). This sketch is from Fig.~2 of reference~\cite{bib17a}.}
\label{fig2}
\end{figure}

In Fig.~\ref{fig2} the energy, electric field and space charge profiles 
across J-I and J-II are represented according to the analysis done 
in~\cite{bib17a}. Because the J-II gap is narrow ($x_g\ll x_{dr}$) and 
the built-in potential is discontinuous (due to the vacuum gap), there 
can be large electric fields there, which can be much greater than in 
the J-I depletion region. As a matter of fact, one can estimate the 
relative magnitude as follows: the J-II electric field is $|\textrm{\bf 
E}_{\textrm{\bf J-II}}|\simeq \frac{V_{bi}}{x_g}$, while the average 
magnitude of the field in J-I is $|\textrm{\bf E}_{\textrm{\bf 
J-I}}|\simeq \frac{V_{bi}}{x_{dr}}$, thus their ratio scales as 
$\frac{|\textrm{\bf E}_{\textrm{\bf J-II}}|}{|\textrm{\bf 
E}_{\textrm{\bf J-I}}|} \sim \frac{x_{dr}}{x_{g}}\gg 1$.
 
Through a mathematical treatment of the device, it has been 
shown~\cite{bib17a,cs} that if some provisos on $x_{g}$ and $x_{dr}$ are 
met, then the electrostatic potential energy in J-II (electrostatic 
energy density times gap volume) is much greater than that in the entire 
depletion region J-I. Furthermore, if the open gap J-II is switched 
closed (thus becoming a second J-I junction), then such an excess energy 
is positively released. Most of the free electronic charges on each gap 
face (see Fig.~\ref{fig2}) disperse through and recombine in the J-II 
bulk.

It is clear that if such a release can be made cyclical through an 
electromechanical nano-apparatus, then this kind of device can exploit 
the thermally driven diffusion across J-I to produce usable work. 
Namely, it appears to violate the Second Law of Thermodynamics in the 
Kelvin-Planck formulation.

In the literature, two kinds of such electromechanical apparatuses have 
been proposed and modeled so far (both analytically and numerically), 
one which uses a Linear Electrostatic Motor (LEM)~\cite{bib17a,cs}, and 
the other using an Hammer and Anvil 
analogue~\cite{bib17b,bib17c,bib19,cs}. The detailed description of 
these interesting devices is beyond the scope of the present paper.

In the following Section we present some arguments (heuristic and 
theoretical) which put the existence of the intense electric field in 
J-II into question. We simply believe that there is no electric field in 
J-II and thus no positive electromechanical energy release is possible 
by switching J-II gap closed.

\section{Some arguments against Solid-State Demon devices}

Now we try to argue that in the above scheme the electric field 
$|\textrm{\bf E}_{\textrm{\bf J-II}}|$ in junction J-II is non-existent.

It is easy to note that the amount of free electronic charge on each gap 
face at J-II depends upon the surface area of those faces. J-II being 
equivalent to a parallel-plate vacuum capacitor, the bigger is the 
surface area $S_{\textrm{\bf face}}$ of the faces, the greater is the 
charge on them, the potential drop being fixed. In our case the 
potential drop is equal to $V_{bi}$, and:

\begin{equation}
Q_{\textrm{\bf J-II\, face}}=CV_{bi}=
\frac{\epsilon_0 S_{\textrm{\bf face}}V_{bi}}{x_g},
\label{eq1}
\end{equation}
where $C$ is the electrostatic capacitance of J-II gap, $\epsilon_0$ is the 
vacuum permittivity, thus the greater is $S_{\textrm{\bf face}}$, the 
greater is $Q_{\textrm{\bf J-II\, face}}$.

Imagine for a moment the following thought experiment. Let us have a 
device similar to that depicted in Fig.~\ref{fig1}, with J-I still not 
closed and with an arbitrarily large surface area of J-II faces. As soon 
as J-I is switched closed, charge diffusion begins and a depletion 
region forms in J-I, together with the built-in potential $V_{bi}$. In 
order to satisfy the {\em path-independence} law and/or the Kirchhoff's 
loop rule, as argued in the cited literature, charges also must start to 
accumulate on each gap face in J-II. This means that a current must 
start to flow through the device bulk and through J-I, until the 
equilibrium is attained. It is easy to see that this current can be made 
arbitrarily high in intensity (if the ohmic resistance $R$ of J-I is 
suitably low) and/or arbitrarily long in duration (high $RC$ time 
constant), since $S_{\textrm{\bf face}}$, and thus $Q_{\textrm{\bf 
J-II\, face}}$, can be arbitrarily large.

\begin{figure}[t]
\begin{center}
\includegraphics[height=12cm]{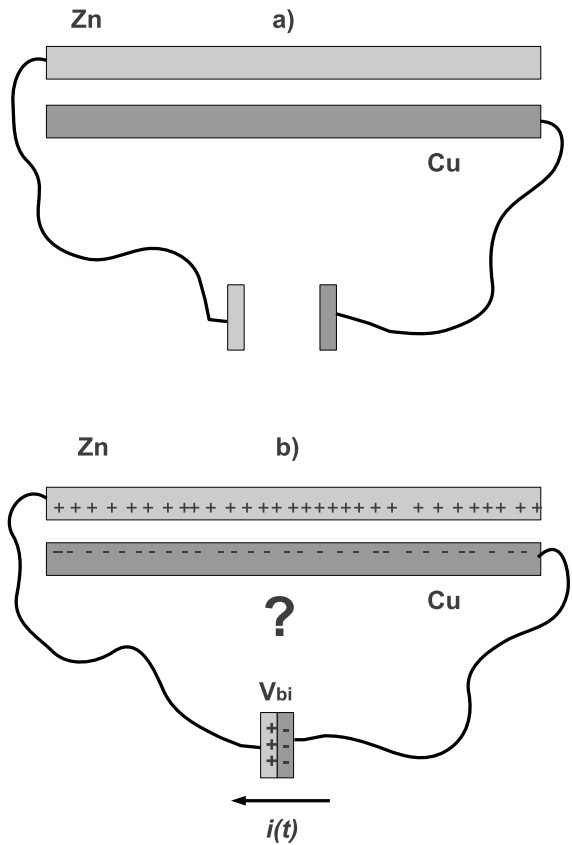}
\end{center}
\caption{Sketch of the first thought experiment described in the text.}
\label{fig3}
\end{figure}

All this is somewhat `unrealistic': J-I can even melt if its section and 
ohmic resistance $R$ are the right ones; or cool down to extremely low 
temperatures, since the energy needed to maintain the current flow 
should come from the thermal agitation in J-I. As a matter of fact, in 
the case in which all the above really happens, the energy stored in the 
parallel-plate equivalent capacitor of J-II gap is 
$\sim\frac{1}{2}CV_{bi}^2$ and it comes exclusively from the thermal 
agitation in J-I. For high values of $C$, the stored electrostatic 
energy becomes huge: with a realistic heat capacity value of junction 
J-I, J-I should cool down very fast and significantly.

A more household analogue can be obtained with two huge metallic plates, 
made of two metals with different work functions. Consider the device 
depicted in Fig.~\ref{fig3}-a. We have two plates, one made of copper 
(Cu) and the other made of zinc (Zn), both placed in vacuum in order to 
eliminate electron exchange with (moist) air and thus avoiding spurious 
charging. They are spatially arranged in order to form a huge 
parallel-plate capacitor. A small wire of Cu starts from the Cu-plate 
and a small wire of Zn starts from the Zn-plate. Both plates are 
initially neutral and not connected to each other through the wires. All 
the system is at a uniform temperature $T$, in order to avoid charge 
accumulation due to the Seebeck/Thomson effects.

\begin{figure}[t]
\begin{center}
\includegraphics[height=7cm]{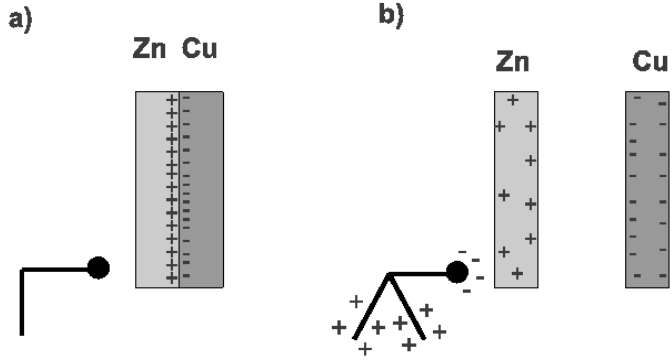}
\end{center}
\caption{Electrostatic behavior (charge diffusion and charge spreading) 
of Cu and Zn plates a) joined and then b) separated 
\cite{kel,cott,io,har}.}
\label{fig4}
\end{figure}

The plate capacitor can be made arbitrarily big, and thus having 
arbitrarily high electrostatic capacitance $C$, since 
$C=\frac{\epsilon_0S}{d}$, where $S$ is the plate surface area and $d$ 
the distance between the plates. As soon as the wire terminals are 
connected, a small Cu-Zn junction forms and a very thin (the junction 
being a metal to metal one) depletion layer is generated along the small 
contact surface (see Fig.~\ref{fig3}-b). The local charge displacement, 
due to diffusion drift, originates a built-in potential $V_{bi}$. If we 
apply the arguments made in \cite{bib17a,bib17b,bib17c,bib19,cs} and 
described in the previous Section, the same potential drop ($V_{bi}$) 
must originate also between the two plates with high electrostatic 
capacitance $C$. This means that an high amount of free charges must 
settle on both plates since, again, $Q=CV_{bi}$. As before, all this 
implies that an arbitrarily high current in intensity and/or arbitrarily 
long in duration must flow through the wires and that the small Cu-Zn 
junction must cool down very fast and significantly. This behavior does 
not match what happens in laboratory experiments and in the real world.

It is already well known that this is not what really happens (see the 
Volta effect~\cite{kel,har}). When two metals with different work 
functions (and similarly, an {\em n-} and a {\em p-}semiconductor) are 
joined, the charge drift is only {\em local} and the charge displacement 
remains localized within the thin depletion layer, in equilibrium. Far 
from the depletion region there is no free charge accumulation. A simple 
laboratory experiment with Cu and Zn plates and a gold-leaf electroscope 
can confirm such a behavior~\cite{kel,cott,io}. Only when the two metals 
are removed apart the charges, initially localized within the depletion 
layer, are free to spread across the surfaces of the metallic 
plates~\cite{cott,io,har}, satisfying electrostatic equipotentiality, 
see Fig.~\ref{fig4}.
  
\begin{figure}[t]
\begin{center}
\includegraphics[height=12cm]{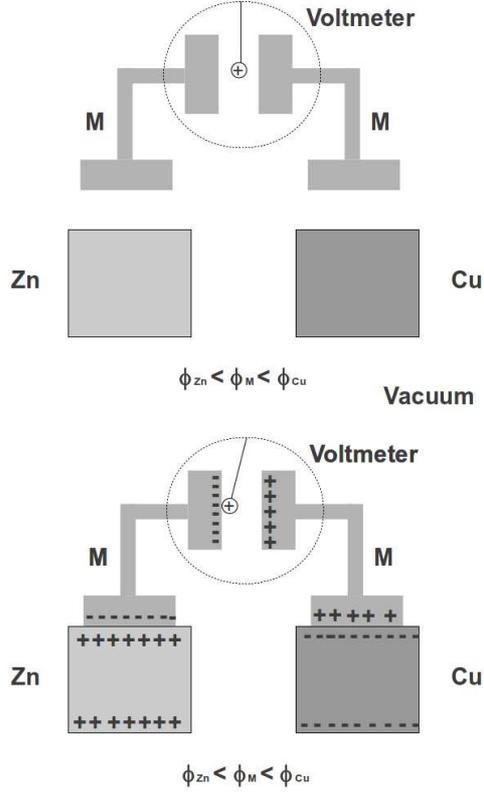}
\end{center}
\caption{Sketch of the second thought experiment described in the text.}
\label{fig5}
\end{figure}

As is written in most introductory textbooks on the subject, the 
difference between work functions of two different materials (metal or 
semiconductor) cannot be measured directly with a normal voltmeter. With 
the thought experiment depicted in Fig.~\ref{fig5} it is easy to show 
that if a diodic or metallic vacuum gap generated and supported a 
capacitive electric field, then it would be possible to measure the 
difference of work functions directly with a normal voltmeter.

As said before, the existence of a capacitive electric field within the 
vacuum gap requires the accumulation of free charges (electrons and 
holes) on the gap faces. Consider now the device sketched in 
Fig~\ref{fig5}. We have three neutral chunks of metal: copper (Cu), zinc 
(Zn) and an unspecified metal (M). As in the previous thought 
experiment, these metals are placed in vacuum in order to eliminate 
electron exchange with (moist) air and thus avoiding spurious charging. 
We choose M such that 
$\phi_{\textrm{Zn}}<\phi_{\textrm{M}}<\phi_{\textrm{Cu}}$, where $\phi$ 
is the work function, as usual. If the contact between metals (or 
semiconductors) with different work functions generated a macroscopic 
charge drift to the opposite (free) sides of the metals (or 
semiconductors), far away from the junction, then the device depicted in 
the Figure would allow the measurement of the difference between 
$\phi_{\textrm{Zn}}$ and $\phi_{\textrm{Cu}}$ directly with a normal 
voltmeter, since we would have opposite free charges on the faces of a 
capacitor made of the same material (M), see Fig.~\ref{fig5}.

Let us now comment the application of the {\em path-independence} law 
and/or Kirchhoff's loop rule. The physical principle at the basis of 
these two laws is the more fundamental law of conservation of energy. 
Conservation of energy demands that a test electronic charge $e$ 
conveyed around a closed path $\gamma$ in the device bulk of 
Fig.~\ref{fig1}, through J-I and J-II at equilibrium, must undergo zero 
net work from {\em all} the forces present along the path. 
Mathematically, we must have,

\begin{equation}
\oint_{\gamma} d\textrm{W}_{ext}=0.
\label{eq2}
\end{equation}

At equilibrium, the only two regions where forces are allowed to be 
non-null are the J-I and J-II regions, as already noted in Section~1. 
When the test charge $e$ crosses J-I, it is subject to the built-in 
electric field force $e\textrm{\bf E}_{bi}$ and to the diffusion force 
$\textrm{\bf F}_{diff}$. We know that at equilibrium $e\textrm{\bf 
E}_{bi}=-\textrm{\bf F}_{diff}$ and that $\textrm{\bf F}_{diff}$ is 
different from zero and constantly present, otherwise $\textrm{\bf 
E}_{bi}$ would soon drop to zero, thus,

\begin{equation}
0=\oint_{\gamma} d\textrm{W}_{ext}=\int_{\textrm{\bf J-I}}
(e\textrm{\bf E}_{bi}+\textrm{\bf F}_{diff})\cdot d{\vec \gamma} + 
\int_{\textrm{\bf J-II}}d\textrm{W}_{ext}=0+\int_{\textrm{\bf J-II}}d\textrm{W}_{ext}.
\label{eq3}
\end{equation}

In the J-II gap there are no diffusion forces, since it is a vacuum gap, 
and eventually we have,

\begin{equation}
0=\int_{\textrm{\bf J-II}}d\textrm{W}_{ext}=\int_{\textrm{\bf J-II}}
e{\textrm{\bf E}_{\textrm{\bf J-II}}}\cdot d{\vec \gamma}=
e|\textrm{\bf E}_{\textrm{\bf J-II}}|x_{g}
\quad\to\quad |\textrm{\bf E}_{\textrm{\bf J-II}}|=0 .
\label{eq4}
\end{equation}

We have presented at least three arguments, the first two more 
heuristic, the third one more theoretical, that suggest that there is no 
electric field within the J-II gap and thus no positive 
electro-mechanical energy release is possible by switching J-II gap 
closed.

\section*{Acknowledgements}
This work has been partially supported by the Italian Space Agency under 
ASI Contract N$^\circ$ 1/015/07/0. The author is grateful to 
Prof.~Daniel P.~Sheehan, Dr.~Assunta Tataranni and Dr.~Gianpietro Summa 
for criticism and insightful discussions. The author is also indebted to 
the three anonymous reviewers for their criticism and suggestions that 
improved the presentation of this article.


\begin{thebibliography}{20}

\bibitem{bib17a} Sheehan, D.~P., Putnam, A.~R., Wright, J.~H.: A 
Solid-State Maxwell Demon. Found.~Phys.~32, 1557-1595 (2002)

\bibitem{bib17b} Sheehan, D.~P., Wright, J.~H., Putnam, A.~R., Pertuu, 
E.~K. Intrinsically biased, resonant NEMS-MEMS oscillator and the second 
law of thermodynamics. Physica~E 29, 87-99 (2005)

\bibitem{bib17c} Sheehan, D.~P., Gross, D.~H.~E.: Extensivity and the 
thermodynamic limit: Why size really does matter. Physica~A 370, 461-482 
(2006)

\bibitem{bib19} Sheehan, D.~P.: Energy, Entropy and the Environment (How 
to Increase the First by Decreasing the Second to Save the Third). 
J.~Sci.~Expl.~22, 459-480 (2008)

\bibitem{cs} \v C\'apek, V., Sheehan, D.~P.: Challenges to the Second 
Law of Thermodynamics--Theory and Experiments, Fundamental Theories of 
Physics, Vol.~146, Springer, Dordrecht, Netherlands, (2005)

\bibitem{kel} L.~Kelvin: On Contact Electricity of Metals. 
Philos.~Mag.~\& J.~Sci., 46, 82-120 (1898)

\bibitem{cott} Cotti, P.: The discovery of the electric current. 
Physica~B 204, 367-369 (1995)

\bibitem{io} D'Abramo, G.: On the exploitability of thermo-charged 
capacitors. Physica~A 390, 482-491 (2011)

\bibitem{har} Harper, W.~R.: The Volta Effect as a Cause of Static 
Electrification. Proc.~R.~Soc.~Lond.~A 205, 83-103 (1951)

\end{thebibliography}
\end{document}